# Probing the coverage of nanoparticles by biomimetic membranes through nanoplasmonics

*Jacopo Cardellini[1,2], Andrea Ridolfi[1,2,3,4], Melissa Donati[1], Valentina Giampietro[1], Mirko Severi[1], Marco Brucale[2,3], Francesco Valle[2,3], Paolo Bergese[2,5,6], Costanza Montis[1,2], Lucrezia Caselli[1,2,7\*], and Debora Berti[1,2\**

1 Department of Chemistry "Ugo Schiff", University of Florence, Florence, Italy
2 CSGI, Consorzio Sistemi a Grande Interfase, University of Florence, Sesto Fiorentino, Italy
3 Istituto per lo Studio dei Materiali Nanostrutturati, Consiglio Nazionale delle Ricerche, 40129 Bologna, Italy
4 Department of Physics and Astronomy and LaserLaB Amsterdam, Vrije Universiteit Amsterdam, Amsterdam, The Netherlands (current affiliation)
5 Department of Molecular and Translational Medicine, University of Brescia, Brescia, Italy
6 Consorzio Interuniversitario Nazionale per la Scienza e la Tecnologia dei Materiali, Florence, Italy
7 Department of Physical Chemistry 1, University of Lund, SE-22100 Lund, Sweden (current affiliation)
*E-mails: debora.berti@unifi.it, lucrezia.caselli@fkem1.lu.se (corresponding authors)

**Abstract**

Although promising for biomedicine, the clinical translation of inorganic nanoparticles (NPs) is limited by low biocompatibility and stability in biological fluids. A common strategy to circumvent this drawback consists in disguising the active inorganic core with a lipid bilayer coating, reminiscent of the structure of the cell membrane to redefine the chemical and biological identity of NPs. While recent reports introduced membrane-coating procedures for NPs, a robust and accessible method to quantify the integrity of the bilayer coverage is not yet available. To fill this gap, we prepared $SiO_2$ nanoparticles ($SiO_2NPs$) with different membrane coverage degrees and monitored their interaction with AuNPs by combining microscopic, scattering, and optical techniques. The membrane-coating on $SiO_2NPs$ induces spontaneous clustering of AuNPs, whose extent depends on the coating integrity. Remarkably, we discovered a linear correlation between the membrane coverage and a spectral descriptor for the AuNPs' plasmonic resonance, spanning a wide range of coating yields. These results provide a fast and cost-effective assay to monitor the compatibilization of NPs with biological environments, essential for bench tests and scale-up. In addition, we introduce a robust and scalable method to prepare $SiO_2NPs/AuNPs$ hybrids through spontaneous self-assembly, with a high-fidelity structural control mediated by a lipid bilayer.

1. Introduction

Over the last decades, numerous efforts have been devoted to the exploitation of the unique properties of inorganic nanoparticles (NPs) for biomedical applications. Despite a large number of NPs developed for biomedical purposes and reports illustrating their in-vitro potential, the route for effective clinical translation is still minimal due to multiple issues[1–5], including poor colloidal stability and limited circulation time in biological fluids, cytotoxic effects, poor targeting ability and uncontrolled accumulation in specific tissues, which eventually leads to low efficacy and unwanted side effects.[6] In a simplistic view, these drawbacks could be considered to be related to a general issue, i.e. the inherent exogenous, synthetic nature of inorganic NPs, and their size, which – being close to that of biomolecules and biological assemblies – can lead to unpredictable behavior when inserted into a biological environment. Coating NPs with a lipid membrane of either synthetic or natural origin is one of the most promising strategies to circumvent this issue, which led to a new class of nanomaterials, i.e., membrane-camouflaged biomimetic NPs. While retaining the physicochemical properties of the core (inorganic) material, the lipid shell of these systems provides biomimetic surface functions, such as immune escape ability and modulation, specific molecular recognition and targeting, enhanced cell adhesion, reduced toxicity, and long circulation time.[7–12] Among the possible sources for membrane camouflaging, synthetic lipid bilayers, whose composition can be conveniently tailored to resemble that of biological membranes, are commonly employed as bioinspired coatings to improve the biocompatibility and pharmacokinetics of NPs,[13,14] as well as to enhance their colloidal stability[15]. Lipid bilayers can be easily functionalized to introduce targeting properties[16,17], steering the carriers towards specific sites, protecting them from the biological environment, and preventing the uncontrolled leakage of drugs.[14] More sophisticated coatings[18,19], employing natural cell membranes (e.g., of red[20–24] and white[25–27] blood cells, cancer cells[28–30], stem cells[31], platelets[32,33], and bacterial walls[34–36]), are currently the focus of intensive research, which has already led to the development of hybrid NPs with superior properties for drug delivery[37–39], in vitro imaging[20,40,41], diagnosis and treatment of cancer[42–44], bacterial infections[45,46] and other diseases[47], anticancer vaccination[48] and detection of viral pathogens[49]. In the palette of natural camouflages, the biomembrane of Extracellular Vesicles' (EVs') represents one of the latest -and more promising- frontiers.[50] EVs are biogenic vesicles naturally secreted by cells, containing lipids, proteins, nucleotides, and metabolites in the inner pool.[51] As compared to other natural membrane coatings, EVs provide unmet targeting abilities, which are connected to their role in cell-cell chemical communication as nano-shuttles for proteins, lipids, and RNA[52,53,54]. Provided by its endogenous origin, the EV membrane also offers near non-immunogenicity, resistance to macrophage uptake[55,53,] and the ability to cross the blood-brain barrier[57,58], as well as enhanced endocytosis efficiency.[59] Coating NPs with the EVs membrane has recently proved as a powerful tool to achieve immune evasion-mediated targeting[60] and selective accumulation at the tumor site[60,61], e.g., through receptor-mediated endocytosis.[62]

Despite the consistent growth of the library of available membrane-coated NPs, achieving complete membrane coverage and developing efficient and reliable methods to quantify its integrity remain significant hurdles. An incomplete membrane coating drastically decreases the colloidal stability of NPs[63–65], and may promote cargo leakage in drug-delivery systems[66]. Moreover, the integrity of the membrane coating modulates the efficiency of macrophage clearance[67] and affects the internalization mechanism of biomimetic NPs[68], as well as their biomedical functions[69–71].

So far, the characterization of the coating integrity has primarily relied on microscopy techniques (e.g., Electron, Confocal Laser Scanning, and Atomic Force microscopies)[72–74], Dynamic Light Scattering (DLS)[72,] and Zeta Potential measurements[75–77]. However, microscopy techniques do not provide ensemble-averaged characterization and generally require specialized equipment and ad-hoc sample preparation. On the other side, scattering-based methods and other traditional approaches (e.g., colloidal stability test in phosphate-buffered saline or fetal bovine serum[68], and sodium dodecyl sulfate-polyacrylamide gel electrophoresis[75,78]) fail in providing a quantitative estimate of the coating extent. More sophisticated techniques, such as mass spectrometry (MS) or liquid chromatography-tandem MS[79–81], only give a rough estimation of the coating degree, i.e., heavily affected by strong assumptions on the morphology and structure of the coating itself.

Here, we report a new colorimetric and spectrophotometric method for quantitatively assessing the membrane coating extent based on the plasmonic properties of citrate-coated gold nanoparticles (AuNPs). AuNPs spontaneously interact with free-standing lipid bilayers, leading to membrane adhesion[82] and AuNPs clustering[83–89]. Recently, we have shown how the spontaneous clustering of AuNPs on synthetic and natural lipid vesicles (such as EVs) can be exploited to gain information on the characteristics of the vesicles themselves, such as their concentration[90], stiffness[91,92], and the presence of protein contaminants[93]. Overall, AuNPs plasmonic properties are emerging as convenient, highly sensitive, and robust probes for lipid interfaces. Given these unique properties, we here test the ability of AuNPs to probe and possibly quantify the membrane coating degree on the surface of inorganic NPs. To this purpose, we prepare biomimetic NPs with an inorganic silica core and a synthetic membrane shell, whose composition mimics the typical one of EVs. Through a combination of structural and spectrophotometric techniques, we investigate the interaction of such membrane-coated SiO$_2$NPs (M-SiO$_2$NPs) with AuNPs, as a function of the membrane coating integrity. Finally, we leverage these findings to estimate the extent of the lipid coverage of M-SiO$_2$NPs, through a simple and fast colorimetric assay.

2. Materials and Methods

*2.1 Materials*

Tetrachloroauric (III) acid (≥99.9%) was provided by Sigma-Aldrich (St. Louis, MO). 1,2-Dioleoyl-sn-glycero-3-phosphocholine (DOPC) (≥99.0%), N-palmitoyl-D-erythro-sphingosylphosphorylcholine (sphingomyelin) (≥98.0%), and cholesterol (≥99.0%) were provided by Avanti Polar Lipids. Sucrose, sodium chloride (NaCl) (≥99.5%), sodium citrate (Na$_3$C$_6$H$_5$O$_7$) (≥99.9%), and calcium chloride (CaCl$_2$) (≥99.999%) were provided by Sigma Aldrich. All chemicals were used as received. Milli-Q-grade water was used in all preparations. Silica Nanoparticles were provided by HiQ-Nano (Arnesano, Lecce, Italy) and are stable in an aqueous buffer and are characterized by a hydrophilic surface with terminal Si-OH functional groups.

*2.2 Preparation of SiO$_2$NPs*

The commercial batch was thoroughly homogenized by vortexing, followed by 30 min bath sonication before use. Subsequently, it was diluted in milliQ water to obtain a final SiO$_2$ concentration of 1.6 mg/mL right before mixing with liposomes.

*2.3 Preparation of liposomes*

To prepare the EVs-mimicking liposomes, the proper amount of DOPC, Sphingomyelin, and cholesterol was dissolved in chloroform (0.87/0.37/1 mol%/mol%), and a lipid film was obtained by evaporating the solvent under a stream of nitrogen and overnight vacuum drying. The film was then swollen and suspended in a warm (40 °C) water solution of sucrose (650 mM), sodium chloride (150 mM), and sodium citrate (10 mM) by vigorous vortex mixing to obtain a final lipid concentration of 7 mg/ml. The resultant multilamellar vesicles (MVL) in water were subjected to 10 freeze−thaw cycles and extruded 10 times through two stacked polycarbonate membranes with a 100 nm pore size at room temperature to obtain unilamellar vesicles (ULV) with a narrow and reproducible size distribution. The filtration was performed with the Extruder (Lipex Biomembranes, Vancouver, Canada) through Nucleopore membranes.

*2.4 Preparation of membrane-coated Silica Nanoparticles (M-SiO$_2$NPs)*

To prepare fully coated SiO$_2$-NPs, 1 mL of a dispersion of uncoated SiO$_2$-NPs in ultrapure water (1.6 mg/mL) was mixed with 1 mL of liposomes dispersion (7 mg/mL) at high stoichiometric excess of liposomes (approximately 1/50 SiO$_2$-NPs/liposomes number ratio), formed in an aqueous environment with high osmolality (650 mM sucrose, 10 mM Sodium Citrate, 150 mM NaCl, 10 mM CaCl$_2$). The formation of the lipid coating on SiO$_2$NPs starts with the adhesion of the vesicles on the SiO$_2$NPs surface due to Van der Waals interactions; this is followed by membrane rupture driven by the transmembrane gradient of osmotic pressure between the inner aqueous pool and the dispersing medium (SiO$_2$NPs were initially dispersed in ultrapure water).

The excess of intact liposomes was then removed through centrifugation (6000 rpm x 30 min each), after which the supernatant was discarded, and the precipitate (containing M-SiO$_2$NPs) collected and redispersed in ultrapure water. This last step was repeated 6 times to fully remove the excess of intact vesicles. To account for possible material loss during the centrifugation cycles, the final concentration of M-SiO$_2$NPs was quantified by ICP-AES (see section S2.3).

To prepare M-SiO$_2$-NPs with different degrees of coverage we employed the same protocol varying the initial SiO$_2$NPs/vesicle ratio. Specifically, 1 mL SiO$_2$-NPs in ultrapure water (1.6 mg/mL) was mixed with 1 mL of liposomes properly diluted to obtain SiO$_2$NPs/vesicle ratios of approximately 1/15, 1/10, 1/5, 1/3, and 1/1. The intact liposomes were removed through centrifugation (6000 rpm x 30 min each), and the precipitate was collected and redispersed in 1 mL of ultrapure water.

*2.5 Synthesis of AuNPs*

Anionic gold nanospheres (AuNPs) of 12 nm in size were synthesized according to the Turkevich−Frens method.[94,95] Briefly, 20 mL of a 1 mM HAuCl4 aqueous solution was brought to the boiling temperature under constant and vigorous magnetic stirring. 2 mL of a 1% citric acid solution were then added to the mixture. The solution was further boiled for 10 min until it acquired a deep red color. The nanoparticle solution was then slowly cooled to room temperature.

*2.6 Cryo-EM*

3 µL of each sample at a SiO$_2$NPs concentration of 1.15 nM were applied on glow-discharged Quantifoil Cu 300 R2/2 grids. The samples were plunge-frozen in liquid ethane using an FEI Vitrobot Mark IV (Thermo Fisher Scientific) instrument. The excess liquid was removed by blotting for 1 s (blot force of 1) using filter papers under 100% humidity and at 10 °C. Cryo-EM data were collected at the Florence Center for Electron Nanoscopy (FloCEN), University of Florence (Italy), on a Glacios (Thermo Fisher Scientific) instrument at 200 kV equipped with a Falcon III detector operated in the counting mode. Images were acquired using EPU software with a physical pixel size of 2.5 Å and a total electron dose of ∼ 50 e−/Å2 per micrograph.

*2.7 Atomic Force Microscopy (AFM)*

NPs were deposited on top of poly-L-lysine (PLL) coated glass coverslips. All reagents were purchased from Sigma-Aldrich Inc (www.sigmaaldrich.com) unless otherwise stated. Menzel Gläser coverslips were cleaned in Piranha solution for 2h and washed in a sonicator bath (Elmasonic Elma S30H) for 30' in acetone, followed by 30' in isopropanol and 30' in ultrapure water (Millipore Simplicity UV). Before each experiment, glass coverslips were treated with air plasma (Pelco Easiglow) and immersed into a 0.01 mg/mL PLL solution in Borate buffer (pH 8.33) at room temperature for 30 minutes. After being thoroughly rinsed with ultrapure water and dried with nitrogen, the coverslips were ready to be used for the AFM experiments. A 10 µl droplet of the SiO$_2$NPs dispersion was deposited on top of the coverslips and left equilibrating for 15 minutes before being inserted into the AFM fluid cell. The concentrations of SiO$_2$NP dispersions were adjusted via trial and error to avoid the formation of NP-clusters, which would ultimately prevent the quantitative determination of their morphology. AFM experiments were performed in PeakForce mode at room temperature on a Bruker Multimode 8 equipped with Nanoscope V electronics, a sealed fluid cell, a type JV piezoelectric scanner, and Bruker SNL10-A probes (with nominal tip radius 2-12 nm and spring constant 0.35 N/m), calibrated according to the thermal noise method[96]. A 50 mM $MgCl_2$, 100 mM KCl solution was used as imaging buffer in order to reduce the electrical double layer (EDL) interaction region between the AFM tip and the NPs[97]. NP height was used to obtain the respective size distributions; given that NPs are spherical rigid objects, their height coincides with the NP diameter and, being unaffected by tip convolution effects, represents a reliable parameter for size estimation.

*2.8 Dynamic Light Scattering (DLS) and ζ-Potential*

DLS measurements at θ = 90° and the ζ-potential determination were performed using a Brookhaven Instrument 90 Plus (Brookhaven, Holtsville, NY). Each measurement was an average of 10 repetitions of 1 min each, and measurements were repeated 10 times. The autocorrelation functions (ACFs) were analyzed through cumulant fitting stopped at the second order for samples characterized by a single monodisperse population, allowing an estimate of the hydrodynamic diameter of particles and the polydispersity index. ζ-potentials were obtained from the electrophoretic mobility u according to the Helmholtz−Smoluchowski equation $\zeta = \left(\frac{\eta}{\varepsilon}\right) \times \mu$ (1), where η is the viscosity of the medium and ε is the dielectric permittivity of the dispersing medium. The ζ-potential values are reported as averages from 10 measurements.

*2.9 Inductively Coupled Plasma - Atomic Emission Spectrometry (ICP-AES)*

The determination of Si and P content in the samples was performed in triplicate by using a Varian 720-ES Inductively Coupled Plasma Atomic Emission Spectrometer (ICP-AES). An accurately measured amount of each sample was diluted to a final volume of 5 mL with 1% suprapure $HNO_3$ obtained by sub-boiling distillation. Each sample was thus spiked with 100 µL of Ge 50 mg/L standard solution used as the internal standard. Calibration standards were prepared by gravimetric serial dilution from a commercial stock standard solution of Si and P at 1000 mg $L^{-1}$. The analytic wavelengths used for Si and P determination were 251.611 and 213.618 nm, respectively, whereas for Ge we used the line at 209.426 nm. The operating conditions were optimized to obtain the maximum signal intensity, and between each sample, a rinse solution constituted of 2% v/v $HNO_3$ was used to avoid memory effects.

*2.10 UV-vis spectroscopy*

UV−vis spectra were recorded with a Cary 3500 UV−vis spectrophotometer. 50 µL of either naked $SiO_2NPs$ or M-$SiO_2NPs$ (at a $SiO_2$ concentration of 1.15 nM with different degrees of coverage were mixed with 300 µL of 6.13 nM AuNPs and incubated for 10 minutes at room temperature in PMMA UV-vis micro cuvettes (maximum volume 1.5 mL). Then, 700 µL of MilliQ water were added to the samples, and after 10 minutes, the spectra were simultaneously recorded with a multiple sample holder in the spectral range 350-800 nm.

*2.11 Small Angle X-ray Scattering*

M-SiO$_2$NPs/AuNPs hybrids were characterized at the SAXS beamline of the synchrotron radiation source Elettra (Trieste, Italy), which was operated at 2 GeV and a 300mA ring current. The experiments were carried out with λ = 1.5 Å, and the SAXS signal was detected with a Pilatus 3 1M detector in the q-range from 0.009 to 0.7 Å$^{-1}$. The SAXS curves were recorded in a glass capillary.

3. Results and Discussion

3.1. Formation of biomimetic membrane-coated SiO$_2$NPs (M-SiO$_2$NPs)

As model inorganic particles, we selected commercial anionic SiO$_2$NPs, given their well-known surface chemistry and the wide range of applications in biomedical research[98,99]. To form the lipid coating, we used DOPC (1,2-dioleoyl-sn-glycero-3-phosphocholine)/Sphingomyelin/Cholesterol (0.87/0.38/1.00 mol%) vesicles, characterized by an average hydrodynamic diameter of 110 nm (PDI 0.150) and a ζ-potential of -15.2 ± 1.3 mV (see section S2 of SI). Composition-wise, these synthetic membranes do not account for the complexity observed in biomembranes, which typically feature an impressive number of different proteins, lipids and (in the case of EVs) nucleic acids. Nevertheless, the above lipid composition has been specifically selected to retain the hallmark feature of EVs membranes, shared among EVs of different sources and biological functions, i.e., the typical enrichment in sphingomyelin and cholesterol, as compared to parental cells.[100] Thus, in spite being highly simplified, these synthetic systems can still represent effective models for the prototypical features of the lipid membrane of EVs. The lipid coating of $SiO_2NPs$ was obtained through a slight modification of a well-established protocol[101] (described in the Materials and Methods section). Briefly, this method relies upon mixing $SiO_2NPs$ in ultrapure water with a high stoichiometric excess of liposomes (≥1/50 $SiO_2NPs$/liposomes number ratio), formed in an aqueous medium with high osmolality. The adhesion of vesicles on the $SiO_2NPs$ surface, driven by Van der Waals attractive forces, is quickly followed by membrane rupture, triggered by the transmembrane gradient of osmotic pressure between the inner aqueous pool of vesicles and the dispersing medium. M-$SiO_2NPs$ were imaged through Cryogenic electron microscopy (Cryo-EM), with Figures 1a and 1b displaying representative Cryo-EM images at different magnifications. While only very few intact vesicles appear to sit onto the $SiO_2$ surface (see Figure 1a, top part, and section S2.1 for additional images), most of $SiO_2NPs$ are either totally or partially (Fig.1b red arrows) surrounded by a nanometric layer, closely following the particle morphology, with an electron density that is intermediate between the ones of $SiO_2$ and the surrounding medium. This layer can be reasonably identified as the bilayer, originally constituting the lipid membrane of vesicles. To gain additional information on the thickness of the surrounding layer, the samples were also imaged by liquid-Atomic Force Microscopy (AFM). Representative AFM images of $SiO_2NPs$ and M-$SiO_2NPs$ are displayed in Figure S2.2, while Figure 1c reports the size distributions obtained for the two samples; the average diameter for the $SiO_2NPs$ is $125 \pm 10$ nm, while the one of M-$SiO_2NPs$ is $140 \pm 15$ nm. The 15 nm difference between the two average diameters is compatible with the presence of a lipid bilayer entirely covering most of the particles of the M-$SiO_2NP$ sample. In addition, we characterized the lipid coating at an ensemble-averaged level, performing DLS and ζ-Potential measurements (Figure 1d). The DLS autocorrelation functions for bare and M-SiO$_2$NPs (Figure 1d) were analyzed through a cumulant fitting stopped at the second order[102]. The inset in Figure 1d summarizes the main results. The hydrodynamic diameter of uncoated SiO$_2$NPs, inferred from the corresponding autocorrelation function, is 165 nm (PDI 0.067). It is worth highlighting that the hydrodynamic size of SiO$_2$NPs is remarkably larger (~40 nm) than the primary particle diameter determined by AFM, which is in line with previous reports[103]. In contrast, the autocorrelation function of M-SiO$_2$NPs decays at longer times, consistent with an increase of hydrodynamic diameter up to 210 nm (PDI 0.1), which is compatible with an extensive formation of a lipid bilayer on the SiO$_2$NPs. Moreover, the characteristic ζ-Potential increases from -40 ± 1 mV (for SiO$_2$NPs) to -22 ± 1 mV (for M-SiO$_2$NPs), which is very close to the one of liposomes (-15 ± 1 mV), further confirmation of the effective lipid coverage of the SiO$_2$NPs' surface. Lastly, we quantified the extent of the SiO$_2$ surface covered by the lipid bilayer employing Inductively Coupled Plasma-Atomic Emission Spectrometry (ICP-AES). This technique allows for exact quantification of the Si and P atoms in the final M-SiO$_2$NPs samples, from which the ratio between the covered and bare SiO$_2$ surfaces can be inferred through simple geometrical models (see section S5.1 in SI). It is worth stressing that these models heavily rely on specific assumptions about the packing of lipid molecules within the membrane formed on SiO$_2$ (see section 2.3 of SI); consequently, the technique can only provide a rough estimation of the degree of coverage of the SiO$_2$NPs surface. Through this approach, a coverage % of 88 ± 8 was estimated for the sample reported in Fig 1.

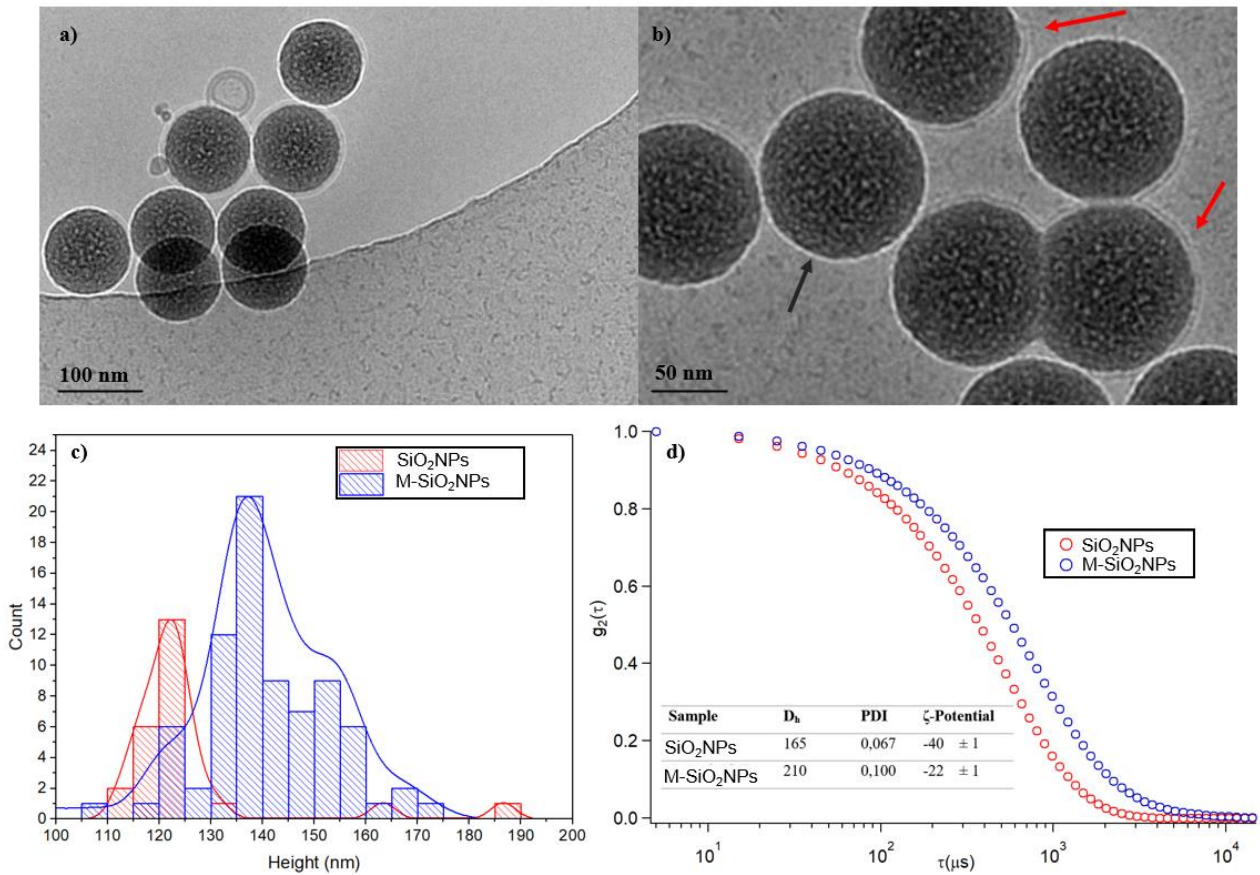

**Figure 1.** a) and b) Cryo-EM images of M-SiO$_2$NPs at different magnifications (red and black arrows identify coated and uncoated areas of M-SiO$_2$NPs, respectively); c) Size distribution of SiO$_2$NPs (red) and M-SiO$_2$NPs (blue) obtained by liquid-AFM imaging; d) Autocorrelation functions of 0.08 mg/mL water dispersion of SiO$_2$NPs (red) and M-SiO$_2$NPs (blue). The inset shows the hydrodynamic diameter (extrapolated by a cumulant fitting) and the ζ-potential values of each sample.

3.2. The interaction of M-SiO$_2$NPs with citrate-coated gold nanoparticles

Once the formation of M-SiO$_2$NPs was demonstrated via different complementary techniques, as discussed in the previous paragraph, we tested the possibility of probing the lipid layer covering the NPs utilizing the plasmonic properties of AuNPs. To this purpose, 1.15 nM bare $SiO_2$NPs (control sample) were first challenged by 6.13 nM Turkevich-Frens AuNPs with an average diameter of 12 nm (PDI 0.095) and -35 ± 3 mV ζ-Potential (see section S3 of SI for AuNPs characterization), for a final $SiO_2$NPs/AuNPs number ratio of ~1/30. The sample was imaged via Cryo-EM (Figure 2a), showing that AuNPs do not interact with bare $SiO_2$NPs, which is expected for the electrostatic repulsion between the two inorganic surfaces, both having a highly negative ζ-Potential. Conversely, when AuNPs are incubated with M-$SiO_2$NPs under the same conditions, a completely different effect is visible: as shown in Figure 2b (see also section S4.1 for additional images), AuNPs spontaneously cluster on M-$SiO_2$NPs, forming AuNPs-decorated M-$SiO_2$NPs composites. This phenomenon, which did not occur with the bare $SiO_2$NPs, is clearly induced by the presence of a lipid bilayer on the $SiO_2$ surface, which mediates the adhesion and clustering of AuNPs on M-$SiO_2$NPs. Cryo-TEM results were complemented by an ensemble-averaged characterization performed through DLS (see section S4.2 of SI), which provided the mean size and polydispersity of $SiO_2$NPs/AuNPs and M-$SiO_2$NPs/AuNPs mixed samples. This characterization showed no interaction between AuNPs and bare $SiO_2$NPs, testified by the presence of two distinct populations within the $SiO_2$NPs/AuNPs sample, whose sizes perfectly match the ones of free AuNPs (~20 nm) and free $SiO_2$NPs (~170 nm). In contrast, a single population of bigger size (~230 nm) was detected for the M-$SiO_2$NPs/AuNPs sample, consistent with the formation of composites in which M-$SiO_2$NPs are decorated by a layer of AuNPs. Remarkably, the clustering of AuNPs on M-$SiO_2$NPs led to an evident color change of the AuNPs dispersion from red to purple/blue (see insets of Figures 2c and 2d) within 10 minutes of incubation, which can be noticed by the naked eye and is connected to a variation of AuNPs' plasmonic properties. Conversely, bare $SiO_2$NPs do not induce significant color variations in the AuNPs' dispersion. The corresponding spectral variations were quantified by means of UV-Vis spectroscopy (results displayed in Figures 2c and 2d). In line with visual observation, the interaction of AuNPs with bare $SiO_2$NPs does not significantly alter the plasmonic features of AuNPs, consisting of the characteristic plasmonic primary peak located at 520 nm (red trace). On the contrary, the interaction with M-$SiO_2$NPs causes a red-shift of the plasmon resonance peak of AuNPs and the occurrence of an additional red-shifted shoulder, a well-established signature of plasmon coupling[91,92], consistent with AuNPs aggregation. As already anticipated, a similar coupling of the plasmons of AuNPs has been observed for AuNPs interacting with natural or synthetic vesicles and has been found to be triggered by AuNPs adhesion to the lipid membrane[82,90], driven by Van der Waals interactions, and promoted by the bending ability of the membrane[90-92]. Such phenomenon has been also conveniently used for determining the concentration, purity, and rigidity of such synthetic or natural vesicles[54,104]. Here for the first time, we show that this phenomenon can also be activated on rigid nanoparticles (such as inorganic $SiO_2$NPs), thanks to the mediating action of a lipid bilayer covering their surface. The clustering of AuNPs on free-standing bilayers was previously found to be promoted by the bending ability of the lipid membrane[88,90–92]; on the contrary, it was expected to be strongly suppressed (or fully prevented) for lipid interfaces formed on rigid supports (e.g., the inorganic core of NPs), where membrane bending ability is strongly reduced. To shed light on this aspect, we directly compared the UV-Vis spectra of AuNPs incubated with M-$SiO_2$NPs with the ones of AuNPs interacting with liposomes presenting the same membrane composition (see section S4.3 of SI); the results showed that the rigid core of $SiO_2$ strongly decreases the aggregation of AuNPs on the lipid membrane (likely due to the increment in the overall stiffness of M-$SiO_2$NPs composites with respect to pristine vesicles), without, however, completely preventing it. In fact, the clustering extent of AuNPs on M-SiO₂NPs is sufficient to detect an apparent color change in AuNPs dispersion and, accordingly, a variation in the corresponding UV-Vis absorbance spectra.

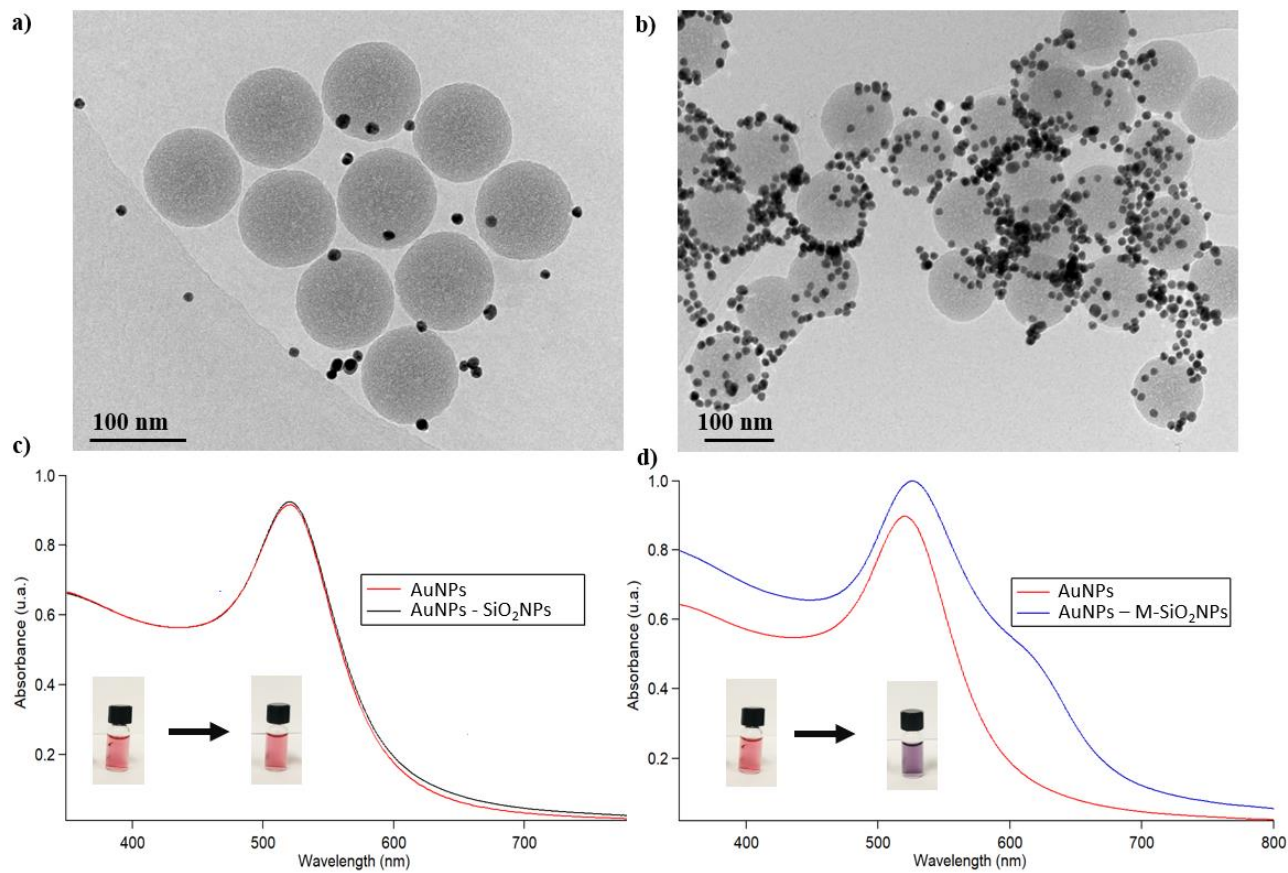

**Figure 2.** Cryo-EM images of (a) SiO₂NPs/AuNPs, (b) M-SiO₂NPs/AuNPs composites, and UV−visible spectra of AuNPs incubated with (c) SiO₂NPs and (d) M-SiO₂NPs. The UV-Vis spectrum of bare AuNPs (red curve) is also reported as a control sample. The visual appearance of AuNPs before and after the incubation with SiO₂NPs and M-SiO₂NPs is reported in the insets of the graphs. See also Figure S11 of SI for UV-Vis control spectra of SiO₂NPs and M-SiO₂NPs in the absence of AuNPs.

3.3 A nanoplasmonic assay to quantify lipid coverage in M-SiO$_2$NPs

Having demonstrated that the self-assembly of citrate-coated AuNPs occurs on lipid-coated SiO$_2$NPs and not on bare ones, we explored how the plasmonic variations of AuNPs are affected by the extent of lipid coverage of the SiO$_2$NPs. To this aim, we applied the same transmembrane osmotic shock-based protocol for synthesizing several SiO$_2$NPs samples with different degrees of coating (see Materials and Methods section). While a NP/vesicle ratio ≥ 1/50 is required to obtain almost full coverage of the silica surface (i.e., ~88%, see previous paragraphs), the coverage extent can be tuned by varying the SiO$_2$NP/vesicle ratio during incubation. Employing SiO$_2$NPs/vesicles ratios of 1/50, 1/15, 1/10, 1/5, 1/3, and 1/1, we realized different hybrids and characterized them through ICP-AES, DLS, and ζ-potential (see section 2.4). Table 1 provides a full overview of such characterization.

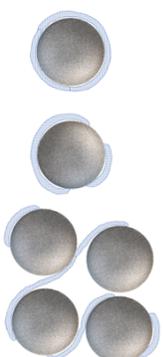

| Sample | $D_h$ (nm) | PDI | ζ-Potential (mV) | SiO$_2$NPs coverage |
|---|---|---|---|---|
| Vesicles | 115 | 0.115 | -11.2 ± 1.1 | / |
| SiO$_2$NPs | 165 | 0.067 | -40.2 ± 0.9 | 0% |
| 1/50 | 210 | 0.100 | -22.7 ± 1.3 | 88 ± 8% |
| 1/15 | 193 | 0.122 | -26.2 ± 1.2 | 68 ± 7% |
| 1/10 | 189 | 0.164 | -28.6 ± 0.7 | 60 ± 6% |
| 1/5 | 193 | 0.152 | -30.3 ± 1.4 | 53 ± 5% |
| 1/3 | 414 | 0.363 | -37.1 ± 3.2 | 33 ± 3% |
| 1/1 | 621 | 0.335 | -36.5 ± 3.5 | 13 ± 1% |

**Table 1.** Hydrodynamic diameter, PDI and ζ-potential values of SiO$_2$NPs, bare vesicles, and M-SiO$_2$NPs obtained using different SiO$_2$NPs-vesicles ratios. The last column reports the SiO$_2$NPs coverage percentage, calculated from the concentration of P and Si measured by ICP-AES for each composition (see SI section S2.3). The inset displays a schematic representation of how the coverage affects the size and stability of the hybrids.

The results obtained from ICP-AES measurements were used for a rough estimation of the coating degrees of SiO$_2$NPs (see SI section S2.3). The coating fraction ranged from 13% to 88% by increasing the number of vesicles employed in the incubation step. The size, colloidal stability, and surface charge of the M-SiO$_2$NPs with different coating degrees were assessed by DLS and ζ-Potential. The ζ-Potential of the M-SiO$_2$NPs decreases in a monotonous trend, passing from -23 ± 1 mV for the 1/50 SiO$_2$NPs/vesicle ratio to -37 ± 4 mV for the 1/1 one, which is very close to the value obtained for bare SiO$_2$NPs. These results are consistent with an increased extent of particle coverage as the number of vesicles per SiO$_2$NP increases. On the other hand, the hydrodynamic size of M-SiO$_2$NPs is stable (around 200 nm) in the range 1/50 to 1/5 SiO$_2$NPs/vesicle ratio, while for higher ratios (i.e., 1/3 and 1/1) the samples display an abrupt increase in the size, reaching very high hydrodynamic diameters (400-600 nm) and polydispersity (around 0.3-0.4). For these dispersions, we observed low colloidal stability, with massive precipitation within 1 h from preparation. This observation can be explained considering that, for high SiO$_2$NPs/vesicle ratios, the coverage on SiO$_2$NPs is only partial. In agreement with some recent reports, a lipid surface coverage lower than 40% induces abrupt precipitation[105] due to the presence of membrane patches on the silica surface, triggering the bridging between different -and partially coated- particles[64,65] (inset in table 1). In the present case, it is reasonable to assume that, for colloidally unstable samples, the vesicles' amount is too low to completely coat the available SiO$_2$ surface, and, therefore, most of the SiO$_2$NPs will present a discontinuous surface coverage. The lipid edges of the bilayer patches represent very high-energy spots, which promote interaction with other uncompleted bilayer shells, inducing particle bridging, and precipitation. In this hypothesis, the dispersions' stability, only achieved for lower $SiO_2NPs$/vesicle ratios (<1/3), can be considered an indirect proof of the formation of intact bilayer shells around the surface of a significant fraction of the $SiO_2NPs$.

We then investigated the interaction of AuNPs with M-$SiO_2NPs$ presenting different coating degrees through UV-vis spectroscopy (Figure 3a). For this purpose, the M-$SiO_2NPs$ were incubated with AuNPs under the same experimental conditions described in section 2.2. The incubation provokes a gradual color variation of the original AuNPs dispersion (see Figure 3a inset), from a ruby red to different shades of purple; this is associated with a broadening of the plasmonic primary signal and the appearance of a red-shifted shoulder, previously also observed for the 1/50 $SiO_2NPs$/vesicles sample (section 2.2). Interestingly, the extent of such variations (especially in terms of the intensity of the red-shifted shoulder) depends on the fraction of membrane-covered $SiO_2NPs$ surface. Specifically, the red-shifted shoulder gets gradually more pronounced with increasing membrane coverage.

To characterize the structure of the AuNP aggregates, we performed Synchrotron Small Angle X-Ray Scattering (SAXS) measurements (Elettra, Trieste, Italy). Figure 3b displays the SAXS profiles obtained for the bare AuNPs and M-$SiO_2NPs$/AuNPs adducts. The scattering signal arises from a form factor $P(q)$, which accounts for the shape and size of the dispersed objects, and from a structure factor $S(q)$, which depends on interparticle interactions. In our experimental conditions, considering the much higher concentration of the AuNPs, the $SiO_2NPs$ contribution to the scattering profiles can be regarded as negligible (see Figure S12 in section S4.4). Therefore, the SAXS profiles shown in Figure 3b are only due to the combination of the P(q) and the S(q) of AuNPs, providing specific information on their structure and aggregation extent. In particular, in the low-q range ($0.082 < nm^{-1} < 0.161$), we notice a power-law dependence between the scattering intensity and q, highlighted by a linear trend in a double-log representation of the profiles. For low scattering vectors values, the scattering signal accounts for the morphology of particles and aggregates. According to the Porod approximation[106], this dependence can be related to the mass fractal dimension of the probed objects, which accounts for the dimensionality of the AuNPs clusters. Specifically, the absolute values of the slopes of the log-log plots can be associated with a Porod coefficient which represents the fractal dimension of the aggregates. The evaluated slope decreases from -0.53 to -1.44 as the membrane coverage increases, as reported in the inset of Figure 3b. In a model-free fashion, such a slope evolution suggests that the aggregation of AuNPs creates larger and more densely packed clusters with increasing $SiO_2NPs$ coverage. Overall, the structural information gained from the SAXS profiles strongly agrees with the plasmonic variations monitored in the UV-Vis spectra.

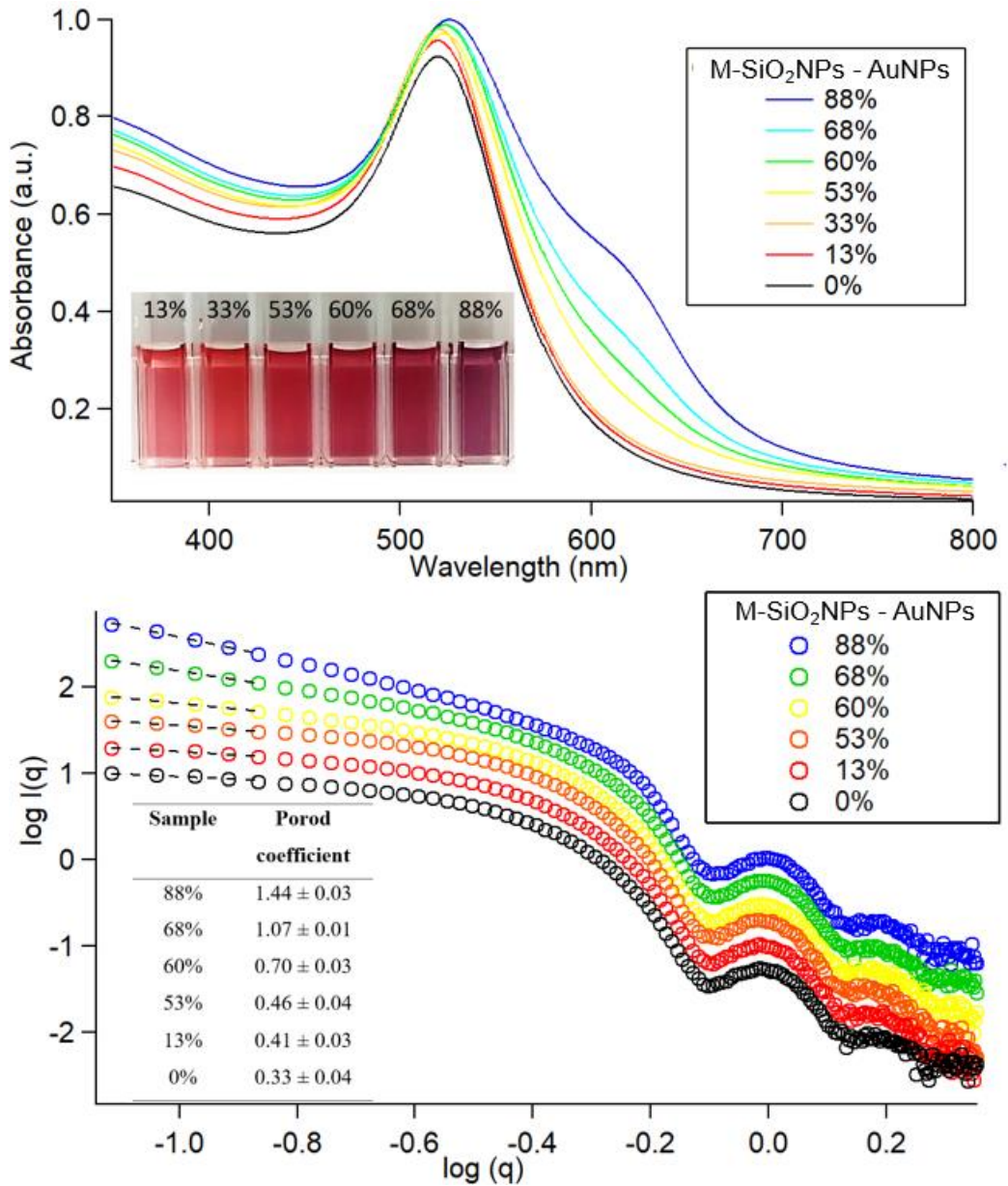

**Figure 3.** (Top) UV−visible spectra of 6.13 nM AuNPs interacting with 1.15 nM M-SiO$_2$NPs with different degrees of coverage, collected after 10 min of incubation; (bottom) SAXS profiles of M-SiO$_2$NPs -AuNPs mixtures with varying degrees of SiO$_2$NPs membrane coverage collected after 10 minutes of incubation. The inset displays the Porod coefficients extracted from the linear fits (dashed black lines) of the double log plots.

The results from SAXS and UV-Vis show that AuNPs clustering on M-SiO$_2$NPs is spontaneous and strictly modulated by the coated surface fraction of SiO$_2$NPs. The plasmonic properties of AuNPs conveniently monitor such dependence through both color and spectral variations of the AuNPs/M-SiO$_2$NPs dispersion. It is then possible to infer the coating degree from a colorimetric assay by introducing a quantitative descriptor of the plasmonic variations of the AuNPs dispersion. To this aim, we selected an optical index recently used to quantify the variation of optical properties of AuNPs dispersions in the presence of synthetic free-standing vesicles[92]. This aggregation index (A.I.) is calculated by dividing the area subtended by the absorbance spectrum in the 560–800 nm range by the total spectral range area (350–800 nm). The results are then normalized for the A.I. of neat AuNPs, so that the A.I. of neat AuNPs' dispersion is equal to 1 and, increasing the particle aggregation, the A.I. value increases. Table 2 summarizes the A.I.s for each M-SiO$_2$NPs/AuNPs sample, where the coating degrees vary.

| SiO$_2$NPs coverage | A.I. |
|---|---|
| 88% | 1.81 ± 0.05 |
| 68% | 1.63 ± 0.04 |
| 60% | 1.45 ± 0.03 |
| 53% | 1.35 ± 0.02 |
| 33% | 1.14 ± 0.01 |
| 13% | 1.07 ± 0.02 |
| 0% | 1 |

**Table 2.** A.I. values obtained for each different membrane coverage of SiO$_2$NPs.

Figure 4 shows the so-determined A.I. values plotted *versus* the membrane coverage % of SiO$_2$NPs (inferred quantitatively by ICP-AES). While the gradual color variation of AuNPs (bottom inset in Figure 4) can already provide some qualitative hints on the degree of membrane coverage, the A.I. and the coverage extent of SiO$_2$NPs are linked by a precise functional relation, which paves the way for developing a spectrophotometric assay for the quantitative determination of membrane coverage. Specifically, the A.I. increases linearly with membrane coverage (r-squared 0.98) over a wide range of coating yields (35-90%). The fitting accuracy decreases when SiO$_2$NPs with a coverage <35% are included in the linear regression, yielding an r-squared of 0.95 (see section 5.2 of SI). This is probably due to the poor colloidal stability of the M-SiO2NPs in low coverage conditions, leading to uncontrolled precipitation of M-SiO$_2$NPs/AuNPs complexes (see Fig. S14 of SI for the correlation between the hydrodynamic size of M-SiO$_2$NPs and the optical response of AuNPs).

While several qualitative methods (e.g., colloidal stability tests[68], and sodium dodecyl sulfate-polyacrylamide gel electrophoresis[75]) can easily detect a very poor (<35%) or absent coverage on NPs, this approach offers a precise -and quantitative- determination of membrane integrity at intermediate and/or high membrane coverage levels, which represents the compositional range of interest for the application of NPs in the biomedical field; as a matter of fact, most of the membrane-related surface functionalities of NPs vanish at low coverages[69–71], while only higher levels of coverage ensure colloidal stability[63,65] of NPs and partial (or complete) preservation of the biological functions[68] of their membrane shell.

Moreover, it is worth noticing that since the aggregation of AuNPs is induced by the presence of the lipid bilayer coating, the assay does not depend on the chemical nature of the nanoparticles' inorganic core and could be extended to NPs with different compositions and physicochemical features.

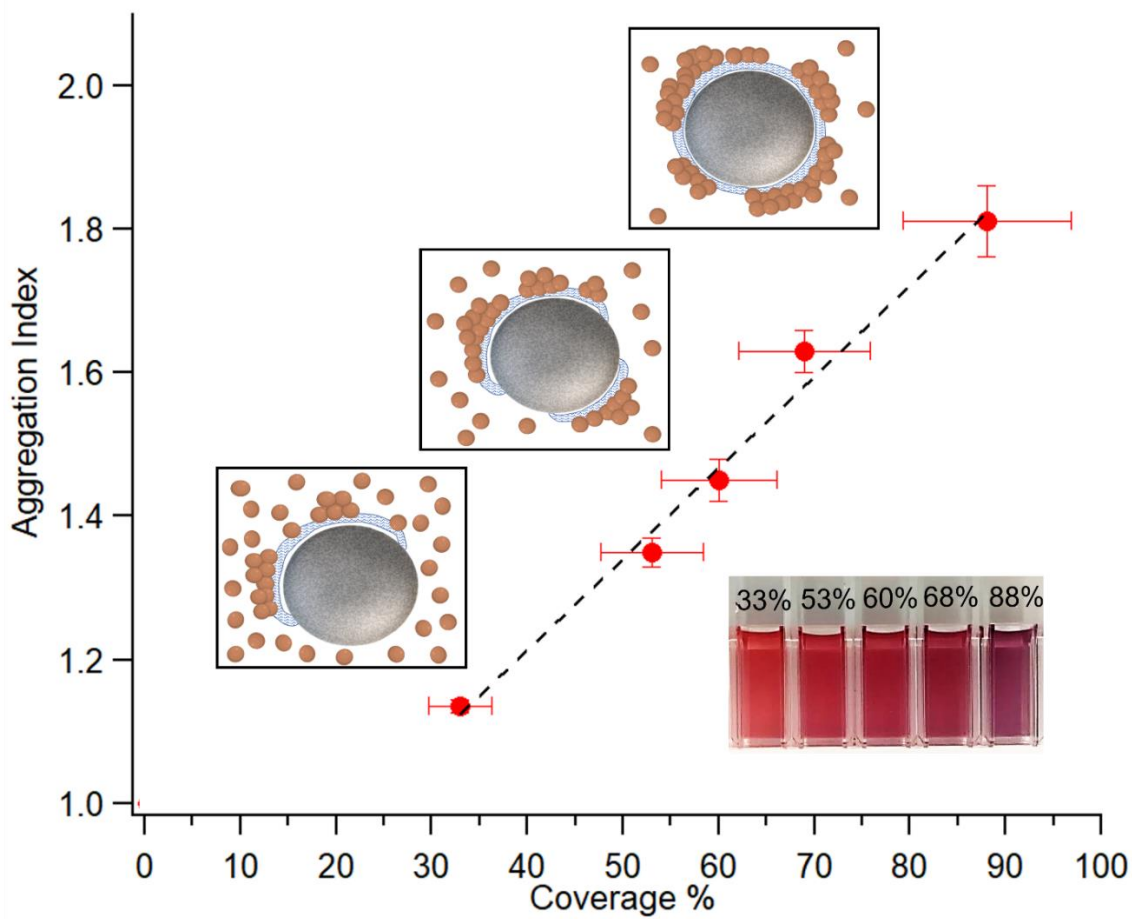

**Figure 4.** A.I. as a function of SiO$_2$NPs membrane coverage, with sketches of AuNPs/M-SiO$_2$ hybrids highlighting how the integrity of the membrane coating affects AuNPs binding and aggregation. Inset: visual appearance of AuNPs incubated with SiO$_2$NPs at different membrane coverages.

4. Conclusions

In the last years, the production of lipid bilayer-coated materials has proved a powerful approach to increase the biocompatibility of inorganic NPs, reduce adverse side effects, and improve their efficiency. To reach this overarching goal, high-fidelity structural control and the availability of analytical methods amenable to scale-up production are critical. Here we described a straightforward and quantitative assay for determining the extent of biomimetic lipid bilayer coverage on inorganic nanoparticles. Based on previous works, we leveraged the plasmonic properties of Turkevich-Frens AuNPs and their sensitivity to the AuNP aggregation extent to develop an effective method for the quantification of lipid membrane coating on inorganic NPs. By challenging $SiO_2$NPs of different -and known- lipid coverage degrees with AuNPs, we show that, in the region of colloidal stability, the plasmonic descriptor of AuNPs' optical properties linearly varies with the amount of lipid coverage and that this dependence can be leveraged for estimating the coverage extent. As a difference from the already available methods[72–74,79–81], such assay provides a fast and high-throughput readout of membrane integrity at an ensemble-averaged level, which only requires cheap reagents, standard lab instrumentation, and limited users experience. From a different -but equally important- perspective, we also showed that a lipid membrane could drive and control the self-assembly of AuNPs on an inorganic nanosized scaffold, which enables the possibility of creating complex hybrid materials composed of an inorganic core, a lipid bilayer shell and a further plasmonic shell of tuneable optical properties. The easiness of preparation, which exploits the spontaneous self-assembly of AuNPs, can inspire the production of multicomponent biocompatible nanomaterials with high structural fidelity and mild experimental conditions.

**Authors information**

**Corresponding Author**

Debora Berti (Department of Chemistry "Ugo Schiff", University of Florence, Florence, Italy and CSGI, Consorzio Sistemi a Grande Interfase, University of Florence, Sesto Fiorentino, Italy, email: debora.berti@unifi.it); Lucrezia Caselli (Department of Physical Chemistry 1, University of Lund, SE-22100 Lund, Sweden, email: lucrezia.caselli@fkem1.lu.se)

**Funding Sources**

This work has been supported by the European Community through the BOW project (H2020-EIC-FETPROACT2019, ID 952183). We also acknowledge MIUR-Italy ("Progetto Dipartimenti di Eccellenza 2018−2022, ref B96C1700020008" allocated to the Department of Chemistry "Ugo Schiff") and Ente Cassa di Risparmio di Firenze for economic support.

**Acknowledgements**

Prof. Paolo Arosio and Karl Normak (Biochemical Engineering Laboratory, ETH Zurich, Switzerland) are acknowledged for sharing their knowledge on the membrane composition and preparation of the liposomal models for EVs employed herein, as well as for fruitful discussion on the design of M-SiO$_2$NPs. The Elettra Synchrotron SAXS facility (Basovizza, Trieste, Italy) is acknowledged for beam time (Proposal id: 20212182). We acknowledge the Florence Center for Electron Nanoscopy (FloCEN) at the University of Florence.

**Abbreviations**

SiO$_2$NPs, silica nanoparticles; AuNPs, gold nanoparticles; M-SiO$_2$NPs, membrane coated silica nanoparticles; EVs, Extracellular Vesicles.

# Supporting Information

Probing the coverage of nanoparticles by biomimetic membranes through nanoplasmonics

*Jacopo Cardellini[1,2], Andrea Ridolfi[1,2,3,4], Melissa Donati[1], Valentina Giampietro[1], Mirko Severi[1], Marco Brucale[2,3], Francesco Valle[2,3], Paolo Bergese[2,5,6], Costanza Montis[1,2], Lucrezia Caselli[1,2,7]\*, and Debora Berti[1,2]\**

1 Department of Chemistry "Ugo Schiff", University of Florence, Florence, Italy
2 CSGI, Consorzio Sistemi a Grande Interfase, University of Florence, Sesto Fiorentino, Italy
3 Istituto per lo Studio dei Materiali Nanostrutturati, Consiglio Nazionale delle Ricerche, 40129 Bologna, Italy
4 Department of Physics and Astronomy and LaserLaB Amsterdam, Vrije Universiteit Amsterdam, Amsterdam, The Netherlands (current affiliation)
5 Department of Molecular and Translational Medicine, University of Brescia, Brescia, Italy
6 Consorzio Interuniversitario Nazionale per la Scienza e la Tecnologia dei Materiali, Florence, Italy
7 Department of Pharmacy, University of Copenhagen, DK-2100 Copenhagen, Denmark (current affiliation)
*E-mails: debora.berti@unifi.it, lucrezia.caselli@fkem1.lu.se (corresponding authors)

**Bibliography**

# 1 Supplementary Characterization of Liposomes

## 1.1 Dynamic Light Scattering and Z-Potential

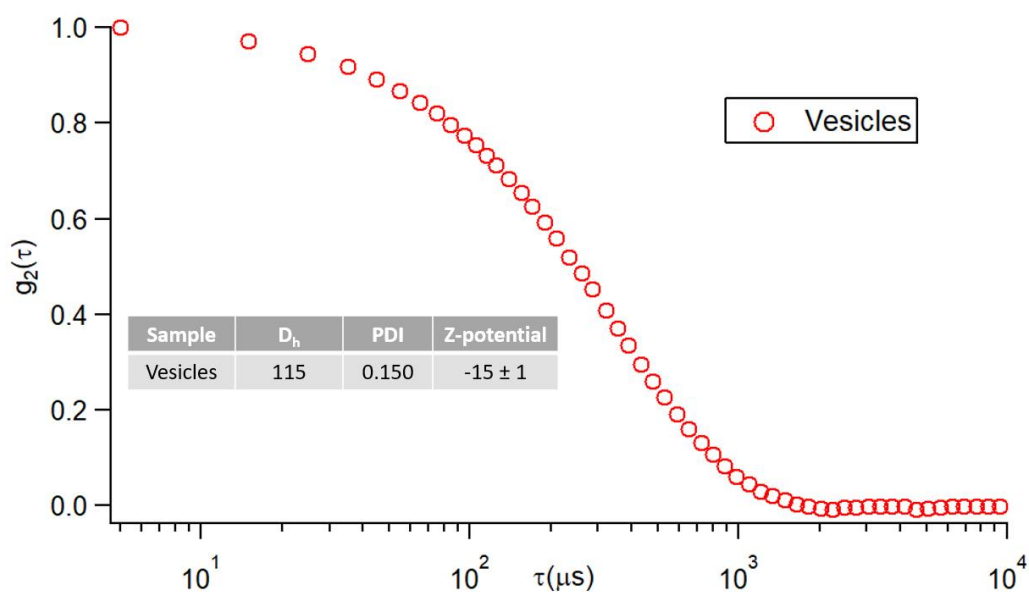

**Fig S1:** Autocorrelation function collected for the vesicles' dispersion at a concentration of 0.35 mg/mL. The table in the inset shows the hydrodynamic diameter and the polydispersity index (PDI), evaluated with the cumulant analysis, and the surface ζ-potential.

## 1.2 Evaluation of Liposomes concentration

The lipid concentration in the starting colloidal dispersion was estimated to be 7 mg/mL from the initial lipid and water amounts employed in the formation and swelling of lipid films, assuming the absence of lipid loss due to the extrusion procedure. The liposomes concentration in the final dispersion was subsequently calculated considering their hydrodynamic diameter (Fig. S1). In particular, the liposomal surface area (surface area=$4\pi r^2$) can be extracted from the liposome diameters; the doubled surface can be subsequently divided by the lipid cross section (approximately 0.5 nm$^2$) in order to obtain the lipid number per liposome, assuming that approximately one half of the lipids is localized in the external leaflet of a liposome, since the bilayer thickness, about 4-5 nm, is negligible with respect to the liposomes' average diameter. Eventually, the total weighted lipid concentration was divided by the total number of lipids per liposome. A liposome concentration of 7.998 x 10$^{13}$ vesicles/mL was obtained, corresponding to a molar concentration of 13 x 10$^{-8}$ M. The liposomal dispersions were diluted to reach a final concentration of 5·10$^{13}$ vesicles/mL before use (corresponding to a molar concentration of 9.5 x 10$^{-8}$ M).

## 2 Supplementary Characterization of M-SiO$_2$NPs

2.1 Cryo-TEM

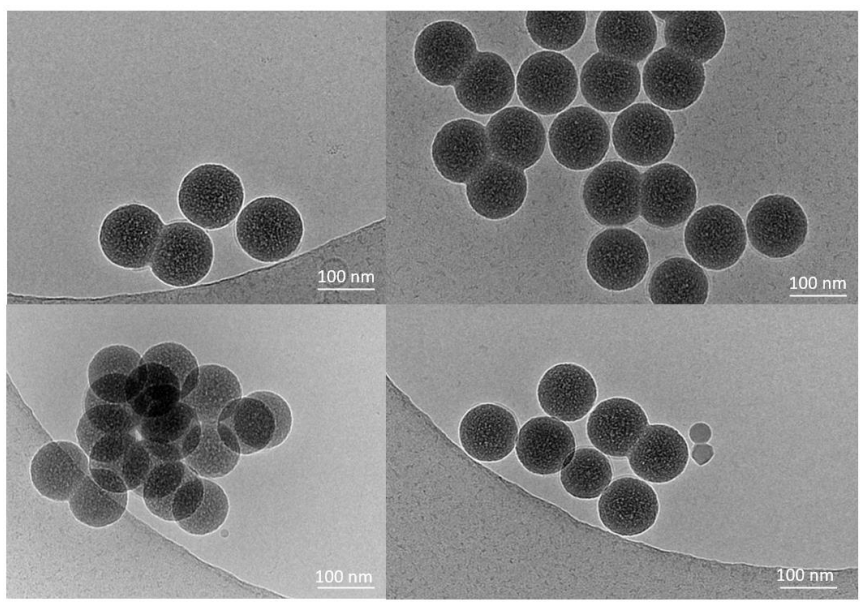

**Fig S2:** Cryo-TEM images of SiO$_2$NPs with a membrane coverage of 88%.

*2.2* Atomic Force Microscopy

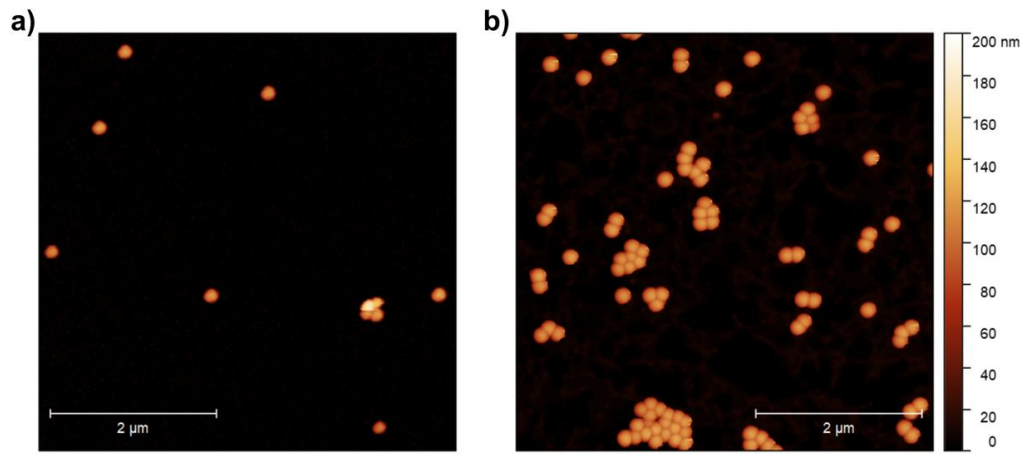

**Fig S3:** Representative AFM images of a) $SiO_2NPs$ and b) $M\text{-}SiO_2NPs$.

2.3 Evaluation of the degree of coverage

To evaluate the degree of lipid coverage of each sample, we performed ICP-AES measurements. With this method we were able to determine (in terms of mg per liters) the amount of P and Si within the dispersion. The obtained values were reported in table S1. Starting from such values we evaluated the lipid coverage of $SiO_2NPs$ for each formulation, according to the following procedure. The amount of P was used for calculating the moles of lipids in the samples. Considering AFM results, the size of $M\text{-}SiO_2NPs$ is approximately 140 nm. Thus, the amount of lipids needed for the formation of a lipid bilayer around a single $SiO_2NP$ can be approximated to be equal to the amount of lipid composing a liposome of 140 nm in diameter (following the procedure reported in section S1.2).

For evaluating the concentration of $SiO_2NPs$ in each sample after 6 cycles of centrifugation, the total mass of $SiO_2NPs$ was obtained starting from the amount of Si measured by the ICP-MS analysis. Then, the mass of a single $SiO_2NP$ was obtained from the density of the $SiO_2NP$ (d=1.9 g/cm$^3$) and the volume of a single $SiO_2NP$ of 120 nm in diameter (904320 nm$^3$). The number of $SiO_2NPs$ present in the sample was obtained by dividing the total mass of $SiO_2NPs$ for the mass of a single $SiO_2NP$. Finally, to obtain the percentage of covered $SiO_2NP$ surface, the total number of lipids (calculated from ICP-AES data as described above) was divided by the number of lipids theoretically needed for fully covering the number of $SiO_2NPs$ present in the sample.

The obtained results are reported in table S3. Each sample was diluted to a $SiO_2$ concentration of 1.15 nм before each further measurement.

| Sample | P mg/L | Si mg/L | SiNPs coverage |
|---|---|---|---|
| **50/1** | 8.9 ± 0.8 | 596.3 | 88 ± 8 % |
| **15/1** | 7.2 ± 0.7 | 626.3 | 68 ± 7 % |
| **10/1** | 5.9 ± 0.6 | 579.0 | 60 ± 6 % |
| **5/1** | 5.0 ± 0.5 | 557.4 | 53 ± 5 % |
| **3/1** | 3.2 ± 0.3 | 570.9 | 33 ± 3 % |
| **1/1** | 1.0 ± 0.1 | 462.3 | 13 ± 1% |

**Table S1:** mg/L concentration of the samples obtained by ICP-AES measurements.

3 Supplementary Characterization of Gold Nanoparticles

3.1 Small Angle X-ray Scattering

SAXS measurements on AuNPs aqueous dispersion were carried out in sealed glass capillaries of 1.5 mm diameter.

The structural parameters (Table S2) of citrate-coated gold nanoparticles were evaluated from the SAXS profile of their diluted water dispersion (2.06 n м) (Figure S1), according to a spherical form factor and a Schulz size distribution. In our concentration range, we can safely assume that there are no interparticle interactions, and that the structure factor S(Q) is equal to 1 in the whole range of scattering vectors. Thus, the scattering profile of the particles derives from their form factor, P(Q). The SAXS spectrum reported in Figure S1 is fully consistent with the characteristic P(Q) of spherical particles with an average diameter of about 5.8 nm. The clear presence of P(Q) oscillations in the high Q region is consistent with a relatively low polydispersity of the synthesized AuNPs.

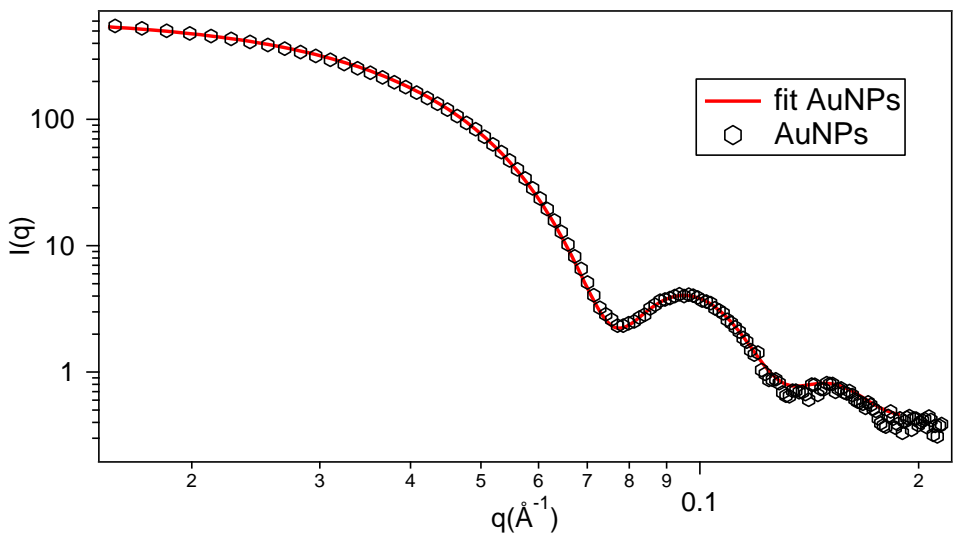

**Figure S4:** Experimental SAXS curve (red markers) obtained for AuNPs, and curve fit (solid black line) according to the Schulz spheres model from the analysis software package SasView. The size and polydispersity obtained from the fitting procedure are summarized in the Table S1 below.

|      | $R_{core}$ (nm) | poly  |
| ---- | --------------- | ----- |
| **AuNP** | 5.8         | 0.095 |

**Table S2:** Structural parameters of the nanoparticles obtained from the analysis of SAXS curves according to the Schulz spheres model.

3.2 Dynamic Light Scattering and ζ-Potential

AuNPs hydrodynamic diameter and surface charge in MilliQ water were evaluated through Dynamic Light Scattering and ζ-Potential, respectively, and reported in inset of figure S5.

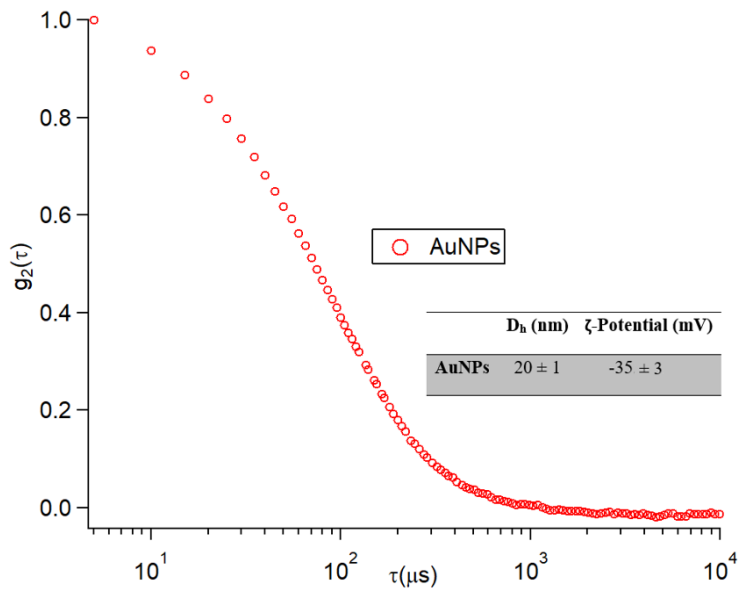

**Figure S5:** Hydrodynamic diameter obtained from Dynamic Light Scattering and surface ζ-Potential of AuNPs.

3.3 UV-vis Spectroscopy

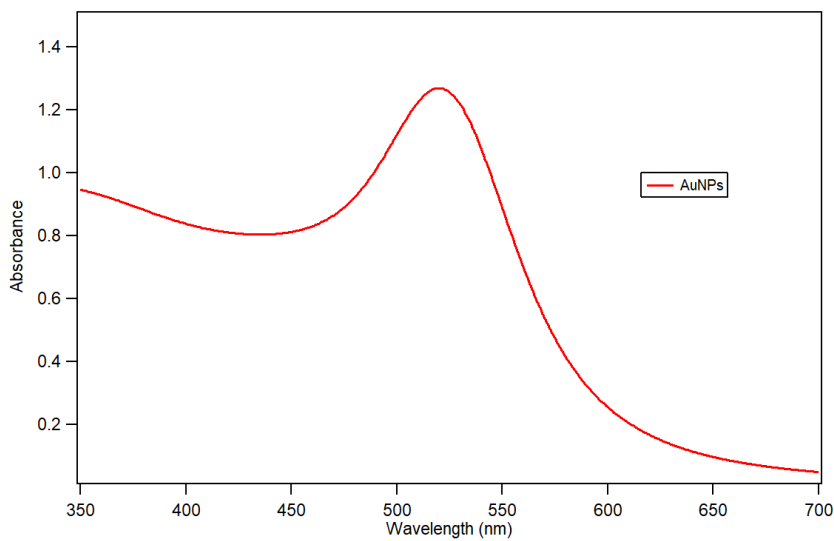

**Figure S6** UV-Vis absorption spectra of AuNPs after 1:3 dilution in water (2.06 nM). The plasmon absorption peak is at around 520 nm.

To further evaluate the AuNPs size through UV-Vis spectroscopy we exploited the following equation[1]:

$$d = \exp\left(B_1 \frac{A_{spr}}{A_{450}} - B_2\right)$$

with $d$ diameter of gold nanoparticles, $A_{spr}$ absorbance at the surface plasma resonance peak, $A_{450}$ absorbance at the wavelength of 450 nm and $B_1$ and $B_2$ are dimensionless parameters, taken as 3 and 2.2, respectively. The diameter value obtained is of 12.3 nm.

The concentration of citrate-coated gold nanoparticles was determined via UV-Vis spectrometry, using the Lambert-Beer law ($E(\lambda) = \varepsilon(\lambda)lc$), taking the extinction values $E(\lambda)$ at the LSPR maximum, i.e. $\lambda = 521$ nm. The extinction coefficient $\varepsilon(\lambda)$ of gold nanoparticles dispersion was determined by the method reported in literature[2], by the following equation:

$$\ln(\varepsilon) = k\ln(d) + a$$

with $d$ core diameter of nanoparticles, and $k$ and $a$ dimensionless parameters ($k = 3{,}32111$ and $a = 10{,}80505$). The arithmetic mean of the sizes obtained by optical and scattering analyses was selected, leading to a $\varepsilon(\lambda)$ of $2.0 \cdot 10^8$ M$^{-1}$cm$^{-1}$. The final concentration of the citrate-coated AuNPs is therefore $\sim 6.13 \cdot 10^{-9}$ M.

## 4 Supplementary Characterization of M-SiO$_2$NPs/AuNPs hybrids

### 4.1 Cryo-TEM

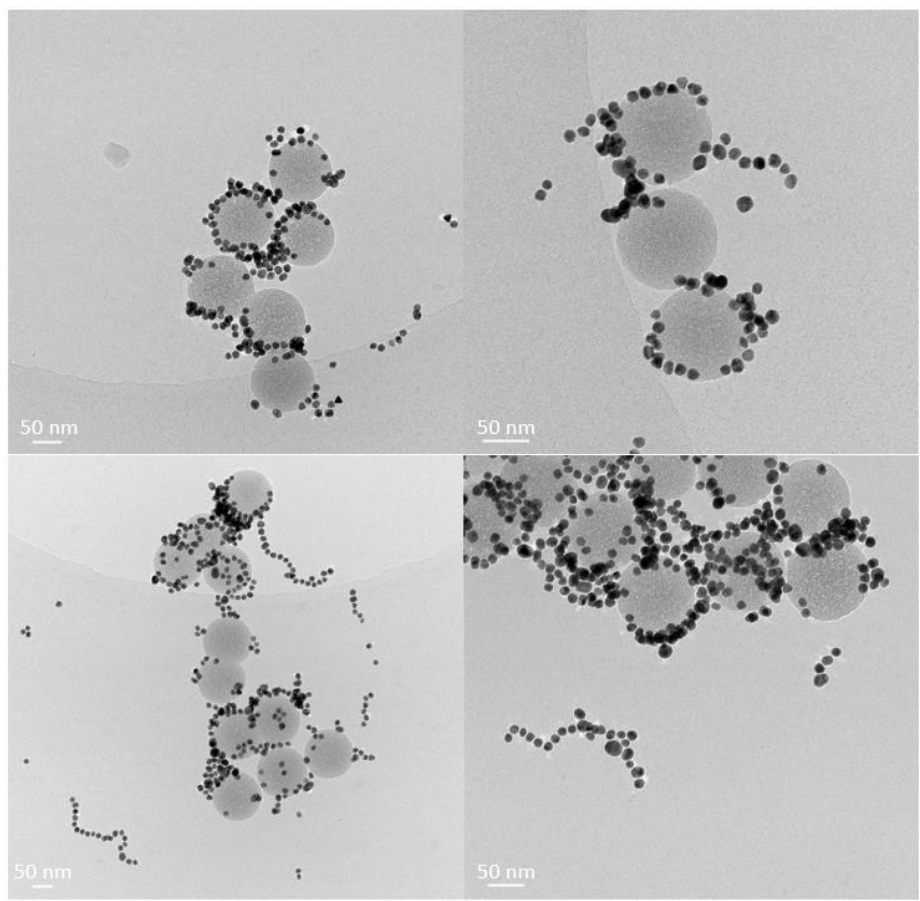

**Figure S7:** Cryo-TEM images of M-SiO$_2$NPs-AuNPs hybrids. M-SiO$_2$NPs are characterized by an almost complete membrane coverage (88%, estimated through ICP-AES).

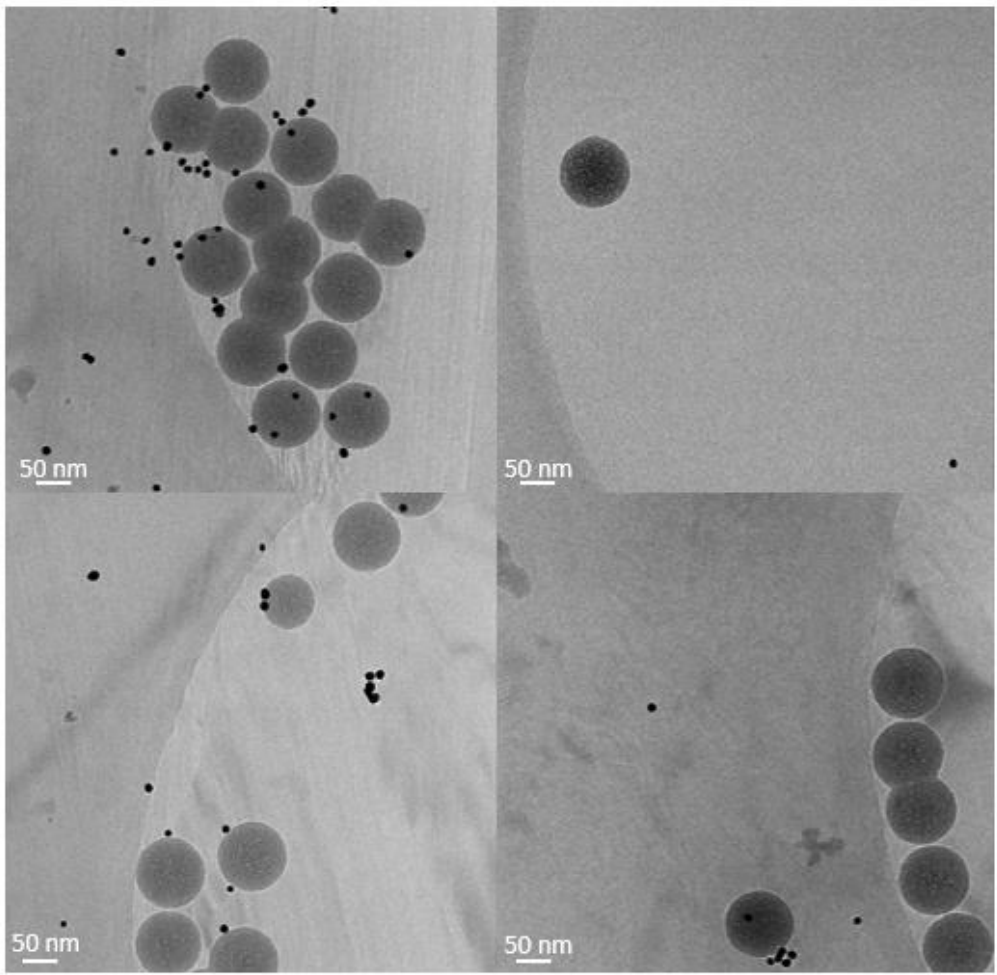

**Figure S8:** Cryo-TEM images of naked SiO$_2$NPs incubated with the AuNPs dispersion.

4.2 Dynamic light scattering

To gain more insights into the aggregation process, we performed DLS measurements on the very same samples used in cryo-TEM experiments. Figure S8 shows the normalized autocorrelation functions measured for the naked SiNPs, the naked SiNPs interacting with AuNPs, and the coated SiNPs interacting with AuNPs. The autocorrelation functions were analysed using a Non-Negatively constrained Least Squares fitting (NNLS). This model, generally used for polydisperse suspensions undergoing Brownian motion, provides a size distribution of the dispersed particles, reported in section figure S8. As shown, the hydrodynamic diameter evaluated for the naked SiNPs is centered at about 170 nm, according to reported cumulant fitting in the main text. The interaction of AuNPs with M-SiO$_2$NPs leads to the formation of larger hybrid objects with a hydrodynamic diameter of about 230 nm. Concerning the naked SiNPs-AuNPs sample, the correlation function is bimodal, suggesting the presence of two separated populations presenting size distributions centred at 20 nm and 170 nm, respectively. Remarkably, the smallest population is consistent with the hydrodynamic dimensions of AuNPs, while the larger one shows very similar sizes to those measured for the uncoated SiNPs. The very fact that DLS measurements are consistent with the presence of two different populations, meaning that the dispersed objects possess very different diffusion coefficients, is a further confirmation that the interaction between AuNPs and SiNPs is mediated by the lipid coating.

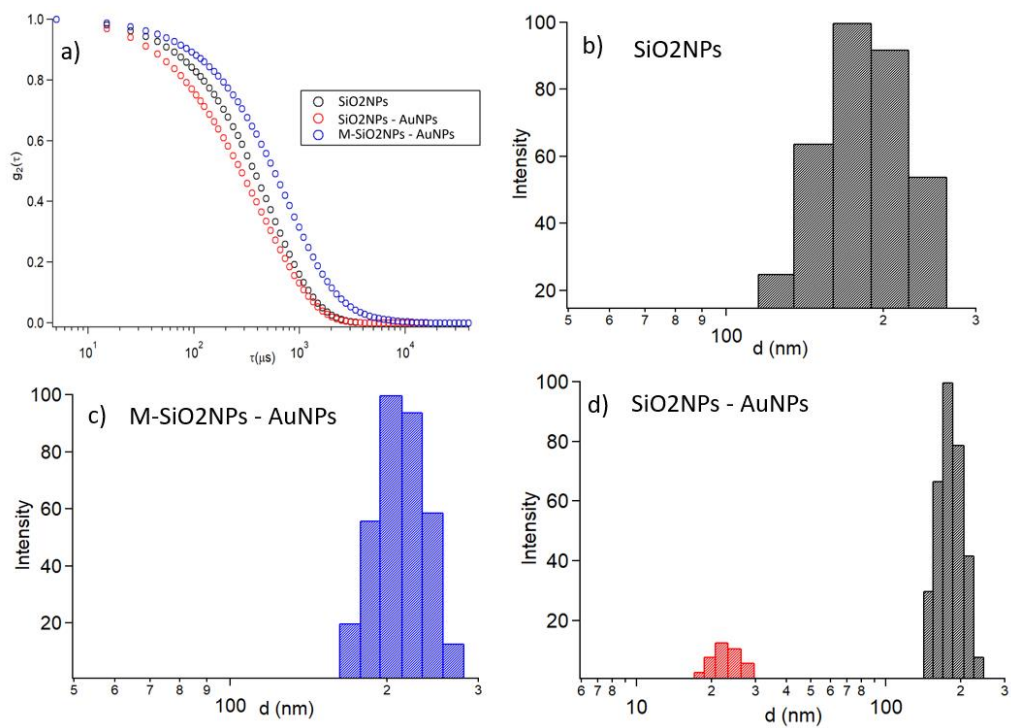

**Figure S9:** a) DLS autocorrelation functions of SiO$_2$NPs, SiO$_2$NPs – AuNPs, , M-SiO$_2$NPs – AuNPs, b) Size distributions obtained from the NNLS fitting of the DLS autocorrelation functions of diluted water dispersion of c) naked SiNPs, b) M-SiO$_2$NPs (with 88% membrane coverage)-AuNPs composites, and d) naked SiNPs-AuNPs mixture.

## 4.3 UV-vis spectroscopy

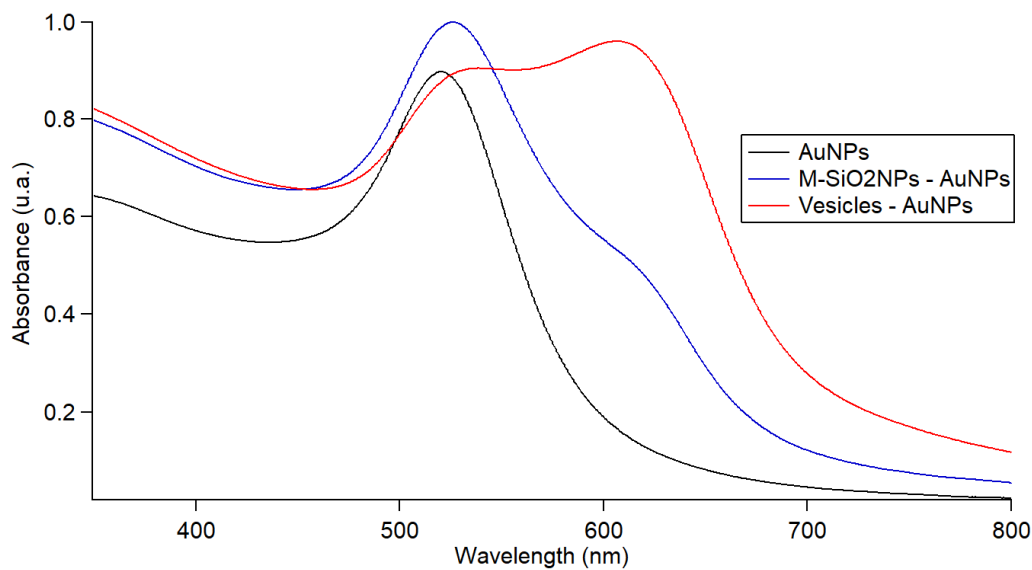

**Figure S10:** UV-vis spectroscopy comparing the plasmonic signal of bare AuNPs and AuNPs incubated with either M-SiO$_2$NPs (with 88% membrane coverage) or free-standing vesicles with the same lipid membrane composition. The concentration of AuNPs was 6.13 nM, while the concentration of vesicles and M-SiO$_2$NPs was 1.15 nM. The plasmonic variation of AuNPs interacting with M-SiO2NPs is significantly less pronounced than the one observed for free-standing vesicles. This indicates a lower AuNPs clustering, likely due to the presence of the rigid silica core (absent in pristine liposomes), increasing the overall stiffness of the composite, in line with recent studies.[3,4]

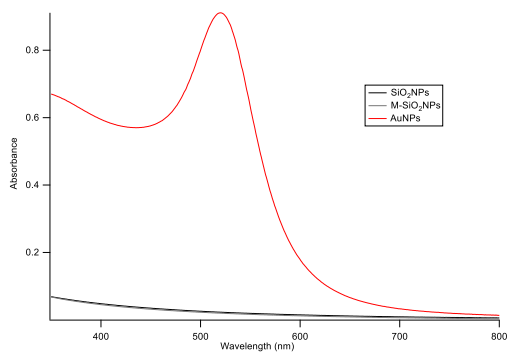

**Figure S11:** UV-vis spectra bare AuNPs (red profile) compared to the ones of SiO$_2$NPs (dark gray) and M-SiO$_2$NPs (with 88% membrane coverage, light gray). The concentration of AuNPs was 6.13 nM, while the concentration of SiO$_2$NPs and M-SiO$_2$NPs was 1.15 nM.

### 4.4 Small-Angle X-Ray Scattering

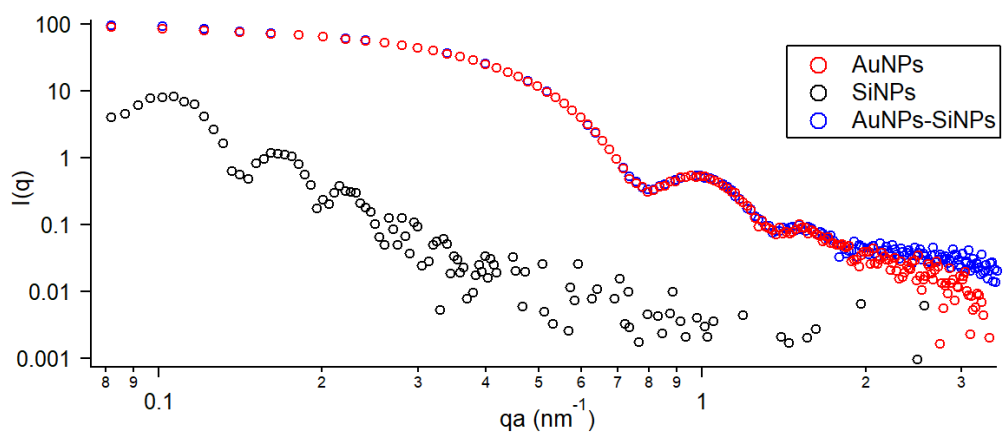

**Figure S12:** Comparison between SAXS profiles of bare AuNPs, SiO$_2$NPs, and SiO$_2$NPs -AuNPs mix obtained by adding 300 μL of 6.13 nM of AuNPs to 50 μL 1.15 nM of SiO$_2$NPs.

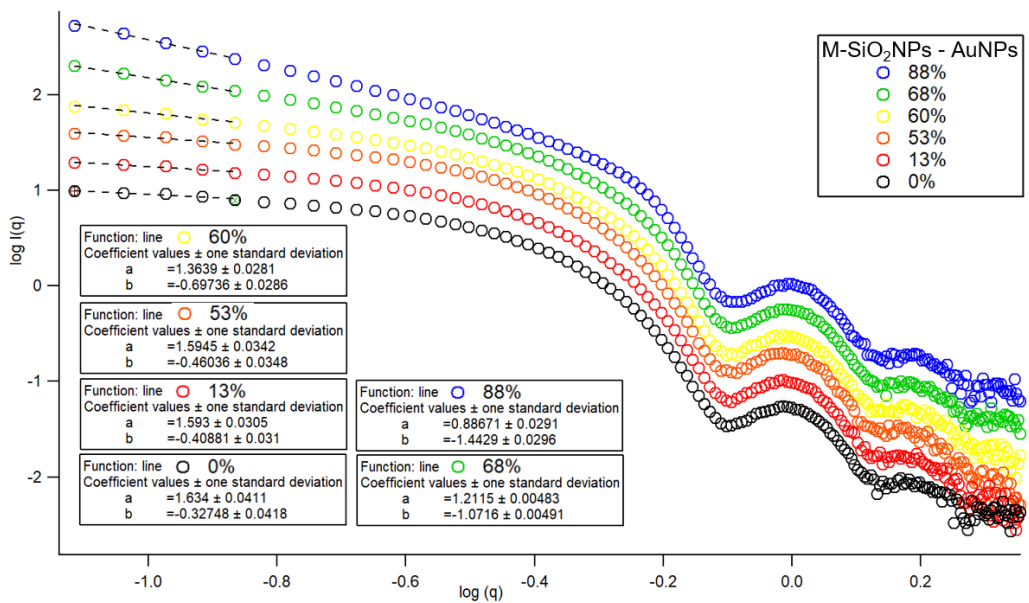

**Figure S13:** Log−Log SAXS profiles of M-SiO$_2$NPs-AuNPs mixtures with different degrees of SiO$_2$NPs membrane coverage collected after 10 minutes of incubation. The dashed lines represent the linear fitting in the Guinier region of the AuNPs (0.082 < nm$^{-1}$ < 0.161). The "b" values in the squared boxes represent the slope values for the different SAXS profiles obtained from the fitting.

| Sample | Porod coefficient |
|---|---|
| **88%** | 1.44 ± 0.03 |
| **68%** | 1.07 ± 0.01 |
| **60%** | 0.70 ± 0.03 |
| **53%** | 0.46 ± 0.04 |
| **13%** | 0.41 ± 0.03 |
| **0%** | 0.33 ± 0.04 |

**Table S3:** Porod coefficient obtained from the slope of the linear fitting of the SAXS profiles in the low q region (Guinier region of the AuNPs 0.082 < nm$^{-1}$ < 0.161).

## 5 Characterization of the M-SiO$_2$NPs/AuNPs as a function of the degree of coverage

### 5.1 Dynamic Light Scattering

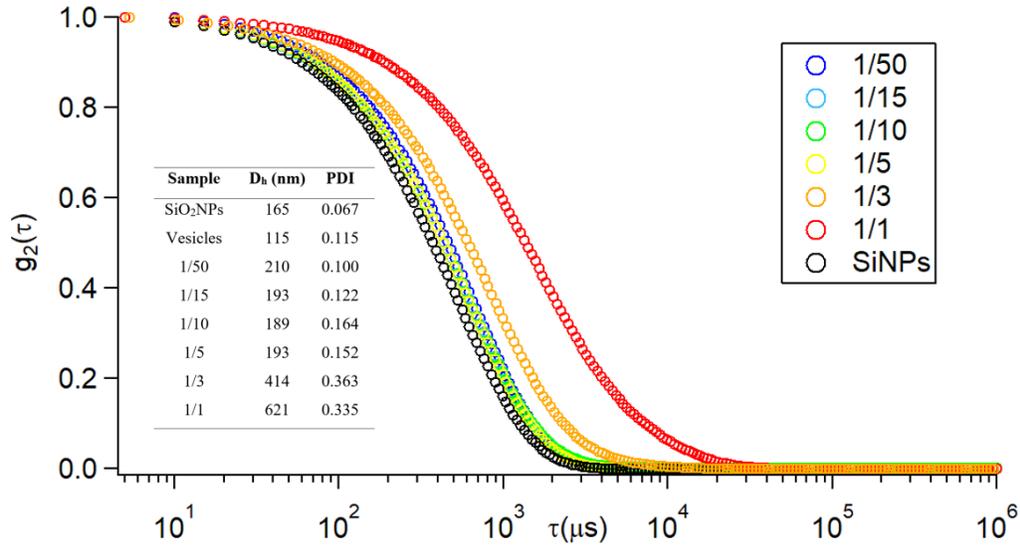

**Figure S14:** Normalized DLS curves obtained for M-SiO$_2$NPs with different coverages. The inset reports the hydrodynamic diameters and PDI of samples, showing the increase in the size of the samples with low coverages.

### 5.2 Nanoplasmonic assay for the quantification of the lipid coverage on SiO$_2$NPs

To set up a colorimetric assay for the estimation of the covered SiNPs (Figure S14), a quantitative descriptor of the plasmonic variations of AuNPs dispersion is required. With this purpose, we selected an aggregation index already used for quantifying the plasmonic variations induced by the presence of synthetic vesicles on the AuNPs dispersion[4]:

$$A.I. = \frac{\frac{A_{560-800}}{A_{350-800}}}{A.I._{AuNPs}}$$

Where $A_{560-800}$ is the area subtended by the absorbance spectrum in the 560-800 nm range, $A_{350-800}$ is the area of the entire spectrum, and $A.I._{AuNPs}$ represents the $\frac{A_{560-800}}{A_{350-800}}$ evaluated for the bare AuNPs dispersion.

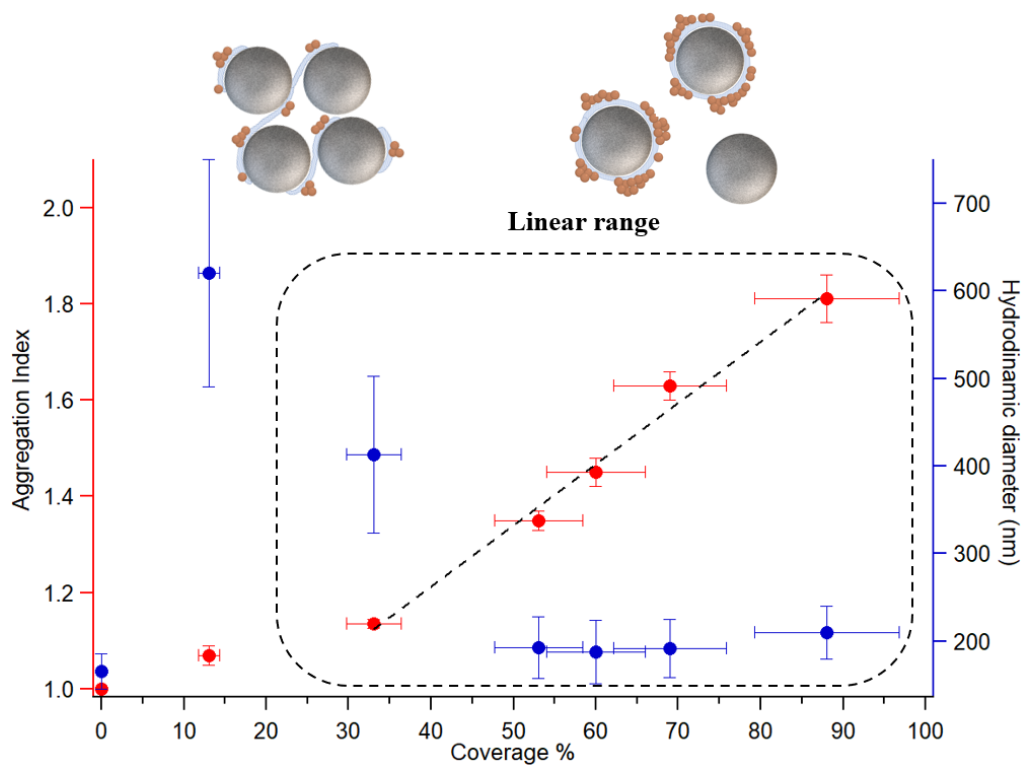

**Figure S15:** A.I. as a function of SiO2NPs coverage. A.I. (left axis) and hydrodynamic diameter (right axis) as a function of SiO2NPs membrane coverage (bottom axis). The linear fit for A.I. vs membrane coverage is also reported, in the 35-100% range of coverage, yielding a r-squared of 0.98.

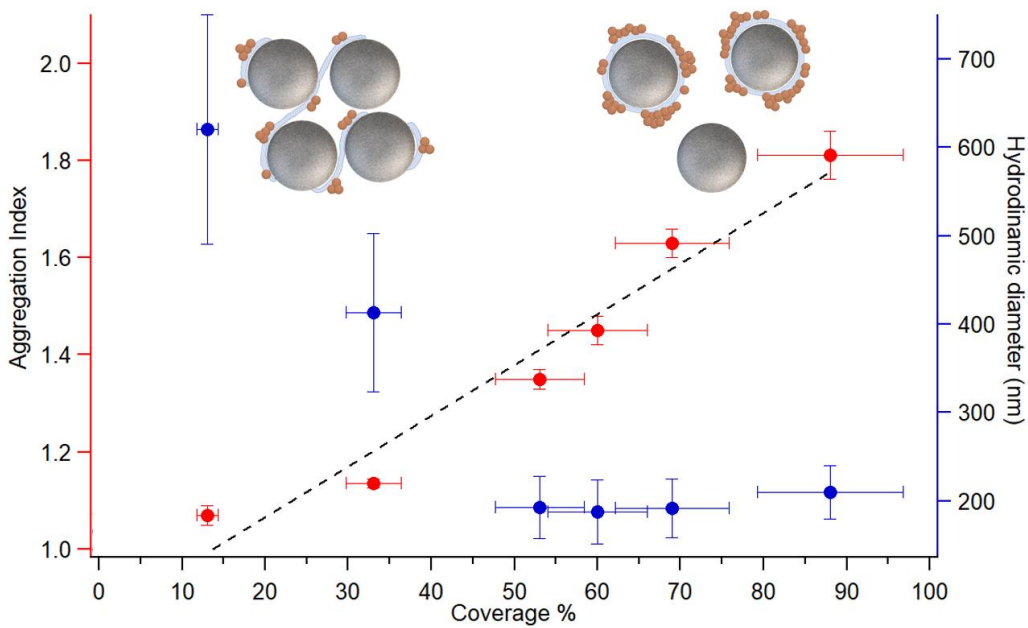

**Figure S16:** A.I. as a function of SiO2NPs coverage. A.I. (left axis) and hydrodynamic diameter (right axis) as a function of SiO2NPs membrane coverage (bottom axis). The linear fit for A.I. vs membrane coverage in the 13-100% range of coverage is reported. The goodness of fit decreases by the inclusion of the A.I. of the SiO$_2$NPs with a coverage of 13% (r-squared of 0.95).

Furthermore, to verify that the linearity of the spectral variation with the lipid coverage of SiNPs is independent of the selected A.I., we further analyzed the spectra with a different aggregation index (A.I. bis). Such A.I.bis was obtained as follow:

$$A.I.bis = \frac{Abs_{max} - Abs_{600}}{\Delta\lambda}$$

Where $Abs_{600}$ is the absorbance evaluated at 600 nm, $Abs_{max}$ is the absorbance of the main peak of the dispersion, and $\Delta\lambda$ is the wavelength difference between 600 nm and the main peak. Each value is then normalized for the A.I. bis obtained from the bare AuNPs spectrum. This A.I. accounts of the spectral variation of the AuNPs spectra considering that the A.I. of bare AuNPs in always equal to 1 and, increasing the particle aggregation, the A.I. value decreases. Table S5 summarized the evaluated A.I.s for each coating extent.

| SiNPs coverage | A.I. |
| --- | --- |
| 88 ± 8 % | 0.70 ± 0.05 |
| 68 ± 7 % | 0.78 ± 0.03 |
| 60 ± 6 % | 0.85 ± 0.03 |
| 53 ± 5 % | 0.91 ± 0.02 |
| 33 ± 3 % | 1 ± 0.01 |
| 13 ± 1% | 0.99 ± 0.02 |
| 0 % | 1 |

**Tab S4:** A.I. bis values obtained for each different membrane coverage of SiNPs.

By plotting the calculated A.I.bis versus the coverage of SiNPs the linear trend of the A.I. with increasing the SiNPs coverage is confirmed for the samples for the lipid coverages higher than approximately 35% (Fig S15).

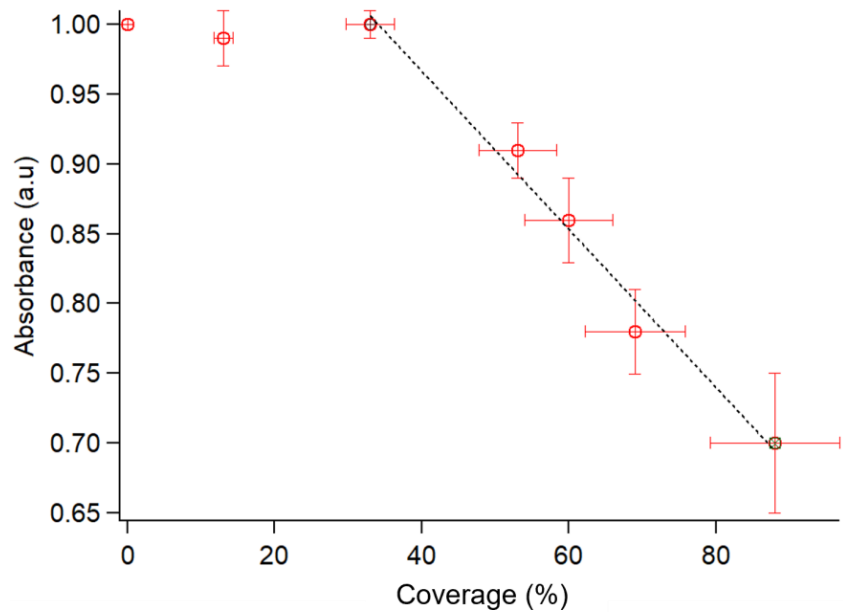

**Figure S17:** Linear trend of the A.I. bis as a function of SiNPs coverage. The samples with low surface coverage (<35%) cannot be fitted by the linear regression.